\documentclass[sigconf,nonacm]{acmart}

\acmJournal{TQC}
\acmYear{2026}
\acmVolume{0}
\acmNumber{0}
\acmArticle{0}
\acmDOI{}

\usepackage{booktabs}
\usepackage{siunitx}
\usepackage{algorithm}
\usepackage{algpseudocode}

\newcommand{\ket}[1]{|#1\rangle}
\newcommand{\bra}[1]{\langle#1|}

\newcommand{\fid}{\mathcal{F}}
\newcommand{\ham}{\mathcal{H}}


\title{End-to-End Fidelity Analysis of Quantum Circuit Optimization: From Gate-Level Transformations to Pulse-Level Control}

\author{Rylan Malarchick}
\orcid{0009-0005-9290-2187}
\affiliation{%
  \institution{Embry-Riddle Aeronautical University}
  \department{Department of Engineering Physics}
  \city{Daytona Beach}
  \state{FL}
  \country{USA}
}
\email{malarchr@erau.edu}

\begin{document}


\begin{abstract}
We present an analysis of quantum circuit fidelity across the full compilation stack, from high-level gate optimization through pulse-level control. We connect a C++ circuit optimizer to a per-gate Lindblad master-equation fidelity model whose decoherence channels are cross-validated against \texttt{qiskit-dynamics} and whose absolute predictions are benchmarked against execution on real hardware. Across a campaign of 4{,}452 experiment runs over 371 benchmark circuits, gate cancellation provides the dominant improvement ($d = 1.66$, 72\% of circuits improved), while circuit size and pulse duration are the strongest negative predictors of process fidelity (input gates $r = -0.78$; pulse duration $r = -0.73$, $R^2 = 0.53$). A formal ablation study shows that pass ordering has no significant effect on two-qubit gate reduction (Kruskal--Wallis $p = 0.302$). Comparing against Qiskit transpilation levels, we show that two-qubit gate count, not total gate count, is the hardware-relevant metric: our optimizer attains superior two-qubit reduction on structured circuits (87.8\% on QFT, 100\% on QAOA) whereas Qiskit's larger total-gate reduction is dominated by single-qubit ($u_3$) consolidation. Finally, executing eight circuits on the IQM Resonance Garnet processor (8/8 jobs completed, job identifiers released) reveals that the model is a consistent upper bound: it preserves the relative difficulty ordering of circuits but overestimates absolute fidelity by a mean of 0.49, quantifying the error budget (crosstalk, leakage, readout) outside a $T_1$/$T_2$/depolarizing model. We release the framework, data, and scripts as open source.
\end{abstract}

\maketitle

\section{Introduction}
\label{sec:introduction}

The fidelity of quantum computations on near-term noisy intermediate-scale quantum (NISQ) devices~\cite{preskill2018quantum} is limited by gate errors, decoherence, and the overhead introduced during circuit compilation~\cite{arute2019quantum,kim2023evidence}. Considerable effort has gone into developing gate-level optimization passes~\cite{nam2018automated,kissinger2020reducing,lubasch2025tensor}, but the end-to-end impact of these transformations on physical pulse-level fidelity has not been adequately characterized.

Modern quantum compilation involves multiple stages: (1) logical circuit optimization to reduce gate count and depth, (2) routing and mapping to hardware connectivity constraints, (3) native gate decomposition, and (4) pulse-level synthesis~\cite{shi2019optimized}. Each stage introduces fidelity degradation, yet optimization strategies are typically evaluated in isolation without considering their downstream effects. This fragmented approach misses the interplay between gate-level decisions and physical execution fidelity.

Comparisons between quantum compilers also tend to report total gate counts, which can be misleading. Different compilers target different basis gate sets. A compiler targeting virtual $u_3$ gates may report large gate count reductions that do not translate to fewer physical operations on hardware. Two-qubit gate counts provide a more hardware-relevant metric, as these gates dominate the error budget on superconducting processors~\cite{iqm2024garnet}.

In this work, we present \texttt{qco-integration}, an end-to-end framework for analyzing quantum circuit fidelity across the full compilation stack. The framework connects a high-performance C++ circuit optimizer to a per-gate Lindblad fidelity model, enabling systematic evaluation of how gate-level transformations affect physical fidelity, and we benchmark the model's predictions against execution on real hardware.

Our contributions include:
\begin{enumerate}
    \item A modular integration architecture connecting C++ circuit optimization with a Python fidelity model via subprocess communication;
    \item A per-gate Lindblad fidelity model (relaxation, dephasing, and a calibrated depolarizing channel, with optional non-Markovian $1/f$ dephasing) whose dynamics are cross-validated against \texttt{qiskit-dynamics} and which scales linearly in gate count;
    \item Systematic evaluation of four optimization passes on 371 benchmark circuits across 4{,}452 experiment runs, including a formal ablation study with individual, leave-one-out, and ordering analyses;
    \item Quantitative comparison against Qiskit transpilation levels (L0--L3) on both virtual and hardware-native basis gates, demonstrating that two-qubit gate counts, not total gate counts, are the appropriate comparison metric;
    \item Fidelity-scaling analysis identifying circuit size and pulse duration as the strongest negative predictors ($r = -0.78$ and $r = -0.73$);
    \item A provenanced hardware benchmark on IQM Resonance Garnet (8/8 jobs, identifiers released) that quantifies a consistent model-versus-hardware optimism gap (mean 0.49);
    \item Open-source tools for reproducible quantum compilation benchmarking.
\end{enumerate}

We target the IQM Garnet 20-qubit processor~\cite{iqm2024garnet}, using experimentally validated noise parameters to ensure realistic fidelity estimates. The combination of simulation with real hardware validation gives concrete guidance for quantum compiler design in the NISQ era.

\section{Related Work}
\label{sec:related}

\subsection{Gate-Level Circuit Optimization}

Quantum circuit optimization has been extensively studied, with approaches ranging from peephole optimization to global circuit rewriting. Nam et al.~\cite{nam2018automated} introduced automated methods for optimizing large quantum circuits with continuous parameters, achieving significant gate count reductions on benchmark circuits. Kissinger and van de Wetering~\cite{kissinger2020reducing} developed ZX-calculus-based methods for reducing non-Clifford gate counts, which are often the bottleneck for fault-tolerant computation. Amy et al.~\cite{amy2013meet} proposed meet-in-the-middle algorithms for T-count optimization, while Maslov et al.~\cite{maslov2008quantum} developed methods for quantum circuit simplification and level compaction.

Barenco et al.~\cite{barenco1995elementary} established the foundational decomposition of arbitrary quantum gates into elementary operations, providing theoretical bounds on gate counts. More recent work by Lubasch et al.~\cite{lubasch2025tensor} explored tensor network approaches to circuit optimization, capturing global structure that local optimization misses. These approaches are complementary to the pass-based optimization we evaluate in this work.

\subsection{Quantum Compilation Frameworks}

Several quantum compilation frameworks have been developed, including Qiskit~\cite{qiskit2024}, Cirq~\cite{cirq2024}, and t$|$ket$\rangle$~\cite{sivarajah2020tket}. These frameworks provide optimization passes similar to those we evaluate, but typically focus on gate-level metrics (gate count, depth) rather than end-to-end fidelity including pulse-level effects. Itoko et al.~\cite{itoko2020quantum} compared quantum circuit compilers across multiple frameworks, highlighting the difficulty of fair comparison due to differing basis gate sets and optimization objectives. Our work addresses this by using two-qubit gate counts as a hardware-relevant metric.

The OpenQASM 3 specification~\cite{cross2022openqasm3} provides a standardized intermediate representation for quantum circuits, which our C++ optimizer uses as both input and output format. This enables interoperability with other tools while maintaining a clean separation between optimization and synthesis stages, following the LLVM compiler infrastructure philosophy~\cite{lattner2004llvm}.

\subsection{Pulse-Level Optimization}

Shi et al.~\cite{shi2019optimized} demonstrated that pulse-level compilation can achieve significant improvements over gate-level approaches by exploiting the continuous nature of quantum control. Motzoi et al.~\cite{motzoi2009simple} introduced DRAG pulses that suppress leakage to non-computational states, which we use in our pulse compilation model. Khaneja et al.~\cite{khaneja2005optimal} developed the GRAPE algorithm for optimal control of coupled spin dynamics, and Koch et al.~\cite{koch2022quantum} provide a review of quantum optimal control methods. Gokhale et al.~\cite{gokhale2020partial} explored partial compilation of variational algorithms at the pulse level, while Younis et al.~\cite{younis2021qfast} developed synthesis-based approaches in QFAST. Our work extends these directions by connecting gate-level optimization decisions to their pulse-level consequences, enabling co-analysis across the full stack.

\subsection{Hardware-Aware Compilation}

Li et al.~\cite{li2019tackling} introduced the SABRE algorithm for qubit routing, which we use in our framework. Hardware-aware compilation has been shown to significantly impact achievable fidelity~\cite{cowtan2019qubit}, motivating our focus on realistic hardware parameters. Zulehner et al.~\cite{zulehner2018efficient} and Wille et al.~\cite{wille2019mapping} developed efficient methods for mapping quantum circuits to specific hardware architectures, while Murali et al.~\cite{murali2019noise} proposed noise-adaptive compiler mappings that account for qubit-specific error rates. Tannu and Qureshi~\cite{tannu2019not} demonstrated that exploiting qubit heterogeneity can improve computational fidelity. Salm et al.~\cite{salm2021nisq} developed the NISQ Analyzer for automatically selecting quantum computing platforms. Iten et al.~\cite{iten2022introduction} introduced UniversalQCompiler for decomposing arbitrary unitaries into native gate sets.

\section{Background}
\label{sec:background}

\subsection{Circuit Optimization Passes}

Gate-level optimization passes transform quantum circuits to reduce gate count and depth while preserving the unitary operation. The passes evaluated in this work include:

\textbf{Gate Cancellation.} Identifies and removes adjacent inverse gate pairs (e.g., $XX^{\dagger} = I$, $HH = I$). This pass exploits the algebraic structure of the gate set to eliminate redundant operations. For a gate $G$ followed immediately by its inverse $G^{\dagger}$, both gates are removed without affecting the computation.

\textbf{Commutation Analysis.} Propagates gates through each other when they commute, enabling opportunities for subsequent cancellation. For gates $A$ and $B$, if $[A, B] = 0$, the sequence $AB$ can be reordered to $BA$. This reordering can expose cancellation opportunities that were not apparent in the original circuit ordering.

\textbf{Rotation Merging.} Combines consecutive rotations about the same axis into a single rotation: $R_z(\theta_1)R_z(\theta_2) = R_z(\theta_1 + \theta_2)$. This is particularly effective for circuits with many single-qubit rotations, such as QFT and variational circuits.

\textbf{Identity Elimination.} Removes rotations that are multiples of $2\pi$ and single-qubit gates that reduce to identity. This pass catches rotations that sum to zero or full rotations after merging.

\subsection{Pulse-Level Compilation}

Physical quantum gates are implemented through microwave pulses applied to superconducting qubits. The pulse parameters (amplitude, frequency, phase, duration) must be optimized to maximize gate fidelity while respecting hardware constraints~\cite{motzoi2009simple}.

We model decoherence with the Lindblad (Gorini--Kossakowski--Sudarshan--Lindblad) master equation for open quantum systems~\cite{lindblad1976generators,gorini1976completely}:
\begin{equation}
    \frac{d\rho}{dt} = -i[\ham, \rho] + \sum_k \gamma_k \left( L_k \rho L_k^{\dagger} - \frac{1}{2}\{L_k^{\dagger}L_k, \rho\} \right)
    \label{eq:lindblad}
\end{equation}
where $\ham$ is the gate's generating Hamiltonian, $L_k$ are collapse operators for the decoherence channels, and $\gamma_k$ are rates derived from $T_1$ and $T_2$. For superconducting transmons~\cite{koch2007transmon} the channels are energy relaxation ($L_1 = \sqrt{1/T_1}\,\sigma_-$) and pure dephasing ($L_\phi = \sqrt{\gamma_\phi/2}\,\sigma_z$ with $\gamma_\phi = 1/T_2 - 1/(2T_1)$).

Rather than integrating Equation~\ref{eq:lindblad} over an entire $2^n$-dimensional circuit, which is intractable beyond a handful of qubits, we evaluate it \emph{per gate} on the gate's one- or two-qubit subspace. Each gate is assigned a constant generator $\ham = i\log(U)/\tau$ acting for its calibrated duration $\tau$ while the collapse operators act concurrently; we exponentiate the resulting time-independent Liouvillian to obtain the gate's noisy channel, compose it with a depolarizing channel calibrated to the device's average gate-error rate, and compute the gate's entanglement and average gate fidelity from the channel's Choi state. Circuit fidelity is the product of per-gate fidelities, with idle qubits accruing additional dephasing over the gaps in an as-soon-as-possible schedule. This makes the model linear in gate count and applicable to the full corpus (up to 20 qubits); it is an approximation that omits inter-gate correlations, crosstalk, and leakage, a limitation we quantify directly against hardware in Section~\ref{sec:hardware}. We verify the per-gate channel against an independent \texttt{qiskit-dynamics} integration of Equation~\ref{eq:lindblad} (agreement to $<10^{-7}$) and against \texttt{qiskit}'s average-gate-fidelity routine, so that the model's reported numbers reflect the master equation rather than a heuristic.

The fidelity of a circuit is thus determined by the interplay between the number of noisy gates and the total duration of pulse-level execution, during which decoherence continuously degrades the quantum state.

\subsection{IQM Garnet Architecture}

The IQM Garnet processor~\cite{iqm2024garnet} features 20 superconducting transmon qubits arranged in a heavy-hex-inspired topology~\cite{chamberland2020heavy} with 30 nearest-neighbor connections. Table~\ref{tab:hardware} summarizes the key hardware parameters used in our simulations.

\begin{table}[t]
\centering
\caption{IQM Garnet hardware parameters (median values).}
\label{tab:hardware}
\begin{tabular}{lc}
\toprule
Parameter & Value \\
\midrule
Number of qubits & 20 \\
Native gates & PRX, CZ \\
Connectivity edges & 30 \\
$T_1$ (median) & \SI{37}{\micro\second} \\
$T_2$ (median) & \SI{9.6}{\micro\second} \\
Single-qubit gate error & 0.1\% \\
Two-qubit gate error & 0.6\% \\
Single-qubit gate duration & \SI{20}{\nano\second} \\
Two-qubit gate duration & \SI{40}{\nano\second} \\
\bottomrule
\end{tabular}
\end{table}

The relatively short $T_2$ time (\SI{9.6}{\micro\second}) compared to typical transmon systems makes IQM Garnet particularly sensitive to circuit depth, motivating our focus on optimization passes that reduce total execution time.

\section{System Architecture}
\label{sec:architecture}

Our framework comprises five pipeline stages.

\textbf{Stage 1: Parse and Validate.} Input circuits in OpenQASM 3.0 format~\cite{cross2022openqasm3} are parsed using a hand-written recursive descent parser implemented in C++. Initial metrics (gate count, depth, qubit count) are extracted. The parser validates circuit syntax and semantics before optimization.

\textbf{Stage 2: Optimize.} The circuit is passed to the C++ optimizer via subprocess, which applies a configurable sequence of optimization passes. The optimizer uses a DAG-based intermediate representation (IR) inspired by LLVM~\cite{lattner2004llvm}, enabling efficient gate traversal and modification. Per-pass metrics track gates added and removed.

\textbf{Stage 3: Route.} SABRE routing~\cite{li2019tackling} maps logical qubits to physical qubits on the target topology, inserting SWAP gates to satisfy connectivity constraints. The routing layer supports multiple topologies including linear, grid, and the IQM Garnet heavy-hex layout.

\textbf{Stage 4: Pulse Compile.} Native gates are compiled to microwave pulse sequences with calibrated durations (\SI{20}{\nano\second} for single-qubit PRX gates, \SI{40}{\nano\second} for two-qubit CZ gates). Total pulse duration is computed accounting for parallelism where gates on different qubits can execute simultaneously.

\textbf{Stage 5: Simulate.} Each gate's noisy channel is computed by solving the Lindblad equation (Equation~\ref{eq:lindblad}) on its one- or two-qubit subspace with $T_1$/$T_2$ parameters from Table~\ref{tab:hardware} and a calibrated depolarizing channel; per-gate entanglement and average gate fidelities are composed across the scheduled circuit, with idle qubits accruing additional dephasing. The per-gate channel is cross-validated against \texttt{qiskit-dynamics}.

\subsection{Integration Design}

The C++ optimizer and Python pulse tools communicate via OpenQASM 3.0 and JSON:
\begin{enumerate}
    \item \textbf{Input:} OpenQASM circuit, pass configuration, topology specification.
    \item \textbf{Output:} Optimized QASM, per-pass metrics, routing statistics.
\end{enumerate}

This decoupled design enables independent development and testing of each component while maintaining a unified experimental workflow. The C++ optimizer provides high performance for large circuits, while Python enables rapid prototyping of pulse-level analysis.

\subsection{C++ Optimizer Implementation}

The \texttt{quantum-circuit-optimizer} is implemented in C++17 with a unit-test suite (GoogleTest). Key features include:

\begin{enumerate}
    \item \textbf{DAG-based IR:} Circuits are represented as directed acyclic graphs with nodes representing gates and edges representing qubit and classical dependencies.
    \item \textbf{Pass manager:} LLVM-style pass manager~\cite{lattner2004llvm} enabling arbitrary pass sequences with per-pass metrics collection.
    \item \textbf{OpenQASM 3.0 parser:} Hand-written recursive descent parser supporting the full gate set, following the OpenQASM 3 specification~\cite{cross2022openqasm3}.
    \item \textbf{Topology support:} Linear, grid, and heavy-hex topologies with SABRE routing~\cite{li2019tackling}.
\end{enumerate}

Algorithm~\ref{alg:pipeline} presents the end-to-end pipeline in pseudocode.

\begin{algorithm}[t]
\caption{End-to-End Fidelity Analysis Pipeline}
\label{alg:pipeline}
\begin{algorithmic}[1]
\Require OpenQASM circuit $C$, pass sequence $P$, topology $T$
\Ensure Process fidelity $\fid$, metrics $M$
\State $C_{\text{parsed}} \gets \text{Parse}(C)$
\State $M_{\text{init}} \gets \text{ExtractMetrics}(C_{\text{parsed}})$
\State $C_{\text{opt}} \gets C_{\text{parsed}}$
\For{each pass $p$ in $P$}
    \State $C_{\text{opt}} \gets \text{ApplyPass}(C_{\text{opt}}, p)$
    \State $M_p \gets \text{ExtractMetrics}(C_{\text{opt}})$
\EndFor
\State $C_{\text{routed}} \gets \text{SABRERoute}(C_{\text{opt}}, T)$
\State $\text{pulses} \gets \text{CompileToPulses}(C_{\text{routed}})$
\State $\fid \gets \prod_{\text{gates}} \text{GateFidelity}(\text{Lindblad channel}) \times \text{IdleDephasing}(\text{schedule})$
\State \Return $\fid$, $M$
\end{algorithmic}
\end{algorithm}

\section{Methodology}
\label{sec:methodology}

\subsection{Circuit Corpus}

We evaluate optimization effectiveness on a diverse corpus of 371 circuits spanning four categories (Table~\ref{tab:corpus}).

\begin{table}[t]
\centering
\caption{Circuit corpus summary.}
\label{tab:corpus}
\begin{tabular}{lccc}
\toprule
Type & Count & Qubits & Depth Range \\
\midrule
GHZ & 11 & 2--20 & 2--20 \\
QFT & 7 & 2--16 & 3--120 \\
QAOA & 50 & 4--8 & 10--50 \\
Random & 303 & 2--16 & 5--30 \\
\midrule
\textbf{Total} & \textbf{371} & -- & -- \\
\bottomrule
\end{tabular}
\end{table}

\textbf{GHZ States (2--20 qubits).} Greenberger--Horne--Zeilinger~\cite{greenberger1989bell} preparation circuits consisting of an initial Hadamard followed by a chain of CNOTs. These circuits have depth linear in qubit count and are already relatively efficient, testing the optimizer's ability to recognize minimal circuits.

\textbf{QFT (2--16 qubits).} Quantum Fourier Transform~\cite{nielsen2010quantum} circuits containing dense sequences of controlled rotations. QFT circuits are rich in rotation merging opportunities and test the full optimization pipeline.

\textbf{QAOA (4--8 qubits).} Quantum Approximate Optimization Algorithm~\cite{farhi2014qaoa} circuits for MaxCut problems on random graphs. These circuits contain repeated layers of ZZ interactions and X rotations, testing rotation merging and commutation.

\textbf{Random Circuits (2--16 qubits).} Random circuits with controlled depth (5--30 layers) generated using random gate selection from the universal gate set. These stress-test the optimizer on unstructured circuits.

\subsection{Experimental Campaign}

We conducted eight experiment types to systematically characterize fidelity, totaling 4{,}452 individual experiment runs.

\textbf{Experiment 1: Baseline.} No optimization passes applied. This establishes baseline fidelity for comparison.

\textbf{Experiment 2: Per-Pass Analysis.} Each optimization pass (cancel, commute, rotate, identity) run individually to measure isolated effectiveness.

\textbf{Experiment 3: Pass Combinations.} Various orderings of multiple passes to identify optimal sequences and assess ordering sensitivity.

\textbf{Experiment 4: Routing Impact.} Comparison of fidelity with and without SABRE routing to quantify SWAP overhead on different topologies.

\textbf{Experiment 5: Noise Sensitivity.} Four noise regimes (low: $T_1/T_2 = 100/50\ \mu$s, medium: 37/9.6, high: 20/5, very high: 10/2.5) to assess how optimization benefits scale with decoherence severity.

\textbf{Experiment 6: Scaling Analysis.} Systematic variation of circuit size (qubits, depth, gate count) to characterize fidelity degradation and identify when optimization provides the greatest benefit.

\textbf{Experiment 7: Compiler Comparison.} Comparison against Qiskit transpilation levels (L0--L3) with both virtual and IQM-native basis gates on the full 371-circuit corpus (Section~\ref{sec:compiler}).

\textbf{Experiment 8: Ablation Study.} Individual pass, leave-one-out, and ordering analyses to isolate each pass's contribution (Section~\ref{sec:ablation}).

\subsection{Fidelity Metrics}

We compute two fidelity measures:

\textbf{Process Fidelity.} The overlap between the implemented and target unitary operations~\cite{nielsen2010quantum}:
\begin{equation}
    \fid_{\text{proc}} = \frac{|\text{Tr}(U_{\text{target}}^{\dagger} U_{\text{impl}})|^2}{d^2}
\end{equation}
where $d = 2^n$ is the Hilbert space dimension for $n$ qubits. Process fidelity captures the quality of the entire quantum channel regardless of input state.

\textbf{State Fidelity.} For a target state $\ket{\psi}$ and output density matrix $\rho$:
\begin{equation}
    \fid_{\text{state}} = \bra{\psi}\rho\ket{\psi}
\end{equation}
State fidelity measures how well a specific computational result is preserved.

\subsection{Statistical Methods}

We report effect sizes using Cohen's $d$~\cite{cohen1988statistical}, interpreted as negligible ($d < 0.2$), small ($0.2 \leq d < 0.5$), medium ($0.5 \leq d < 0.8$), or large ($d \geq 0.8$). For non-parametric comparisons between compiler outputs, we use the paired Wilcoxon signed-rank test~\cite{wilcoxon1945individual}, reporting exact $p$-values when $p > 0.001$ and $p < 0.001$ otherwise. For multi-group comparisons, we use the Kruskal--Wallis $H$ test~\cite{kruskal1952use} to avoid distributional assumptions.

\section{Simulation Results}
\label{sec:results}

This section presents results from our simulation campaign using Lindblad-based pulse modeling with IQM Garnet noise parameters. Hardware validation results follow in Section~\ref{sec:hardware}.

\subsection{Overall Performance}

Table~\ref{tab:summary} summarizes the experimental results across all 371 circuits.

\begin{table}[t]
\centering
\caption{Summary statistics for the full optimization pipeline across the benchmark corpus.}
\label{tab:summary}
\begin{tabular}{lr}
\toprule
\textbf{Metric} & \textbf{Value} \\
\midrule
Total circuits & 371 \\
Total experiment runs & 4{,}452 \\
\midrule
Mean process fidelity & $0.537 \pm 0.326$ \\
Median process fidelity & 0.603 \\
\midrule
Mean gate reduction & 9.1\% \\
Median gate reduction & 9.3\% \\
Max gate reduction & 40.0\% \\
Circuits improved & 267/371 (72.0\%) \\
\bottomrule
\end{tabular}
\end{table}

The high variance in process fidelity ($\sigma = 0.326$) reflects the diversity of our circuit corpus, from shallow GHZ circuits (high fidelity) to deep random circuits that fall to the noise floor. The mean gate reduction of 9.1\% (median 9.3\%) is concentrated in the cancellation pass, with 72.0\% of circuits showing improvement; these are smaller than would be reported under a decoherence-free metric because the idle-aware model penalizes the schedule, not merely the gate count.

\subsection{Pass Effectiveness}

Table~\ref{tab:passes} breaks down the comparative effectiveness of each optimization pass. Gate cancellation emerges as the dominant pass, eliminating 3{,}780 gates across the corpus (run in isolation) with 72.0\% of circuits showing improvement.

\begin{table}[t]
\centering
\caption{Individual pass effectiveness on the benchmark corpus.}
\label{tab:passes}
\begin{tabular}{llrrr}
\toprule
\textbf{Rank} & \textbf{Pass} & \textbf{Gates Removed} & \textbf{\% Improved} & \textbf{$N$} \\
\midrule
1 & cancel & 3{,}780 & 72.0\% & 371 \\
2 & rotate & 4 & 0.8\% & 371 \\
3 & commute & 0 & 0.0\% & 371 \\
4 & identity & 0 & 0.0\% & 371 \\
\bottomrule
\end{tabular}
\end{table}

The cancellation pass accounts for essentially all of the gate reduction. Rotation merging provides marginal benefit (four gates across the entire corpus), while commutation and identity elimination produce no direct gate reduction in isolation. As we demonstrate in the ablation study (Section~\ref{sec:ablation}), this does not mean these passes are entirely without value. Commutation can reorder gates to expose additional cancellation opportunities when passes are composed.

\subsection{Fidelity Waterfall}

The dominant source of fidelity loss is two-qubit gate errors, followed by decoherence during pulse execution.

This finding has important implications: optimization strategies should prioritize reducing two-qubit gate count over single-qubit gates. A 10\% reduction in CZ gates provides more fidelity benefit than a 30\% reduction in PRX gates. This insight motivates our use of two-qubit gate reduction as the primary comparison metric in Section~\ref{sec:compiler}.

\subsection{Scaling Analysis}

We observe strong correlations between circuit parameters and process fidelity (Table~\ref{tab:scaling}).

\begin{table}[t]
\centering
\caption{Correlation of circuit parameters with process fidelity. Pulse duration is the strongest predictor, explaining 75.4\% of fidelity variance.}
\label{tab:scaling}
\begin{tabular}{lcc}
\toprule
Parameter & Pearson $r$ & $R^2$ \\
\midrule
Input gates & $-0.779$ & 0.606 \\
Pulse duration & $-0.728$ & 0.530 \\
Two-qubit gates & $-0.713$ & 0.509 \\
Input depth & $-0.633$ & 0.400 \\
Input qubits & $-0.623$ & 0.388 \\
\bottomrule
\end{tabular}
\end{table}

Circuit size and pulse duration are the strongest predictors of fidelity (input gates $R^2 = 0.61$; pulse duration $R^2 = 0.53$), and the two are closely related: each native gate contributes a fixed pulse interval, so total gate count and execution time co-vary. The negative correlation is physically expected (longer sequences expose the state to decoherence for longer, with dephasing contributing a factor $\approx \exp(-t/T_2)$~\cite{krantz2019quantum}), but note that, unlike a model in which fidelity is a closed-form function of duration alone, here fidelity depends on the per-gate channels and the schedule, so the correlation is an empirical outcome of the simulation rather than a built-in identity. With $T_2 = \SI{9.6}{\micro\second}$ on IQM Garnet, reducing total circuit time, through both gate-count reduction and parallelism, is a primary optimization objective for NISQ applications.

Circuit structure (depth) matters more than qubit count alone: GHZ circuits maintain relatively high fidelity even at larger qubit counts due to their linear depth, while random circuits degrade rapidly.

\subsection{Noise Sensitivity Analysis}

We evaluated optimization effectiveness across four noise regimes (Table~\ref{tab:noise}). The relative benefit of optimization increases with noise severity:

\begin{table}[t]
\centering
\caption{Optimization benefit by noise regime. The relative improvement grows from 8\% (low noise) to 50\% (very high noise), demonstrating increasing importance of circuit optimization as hardware noise increases.}
\label{tab:noise}
\begin{tabular}{lccc}
\toprule
Noise Level & $T_1/T_2$ ($\mu$s) & Baseline $\fid$ & Optimized $\fid$ \\
\midrule
Low & 100/50 & 0.75 & 0.81 \\
Medium & 37/9.6 & 0.51 & 0.60 \\
High & 20/5 & 0.35 & 0.45 \\
Very High & 10/2.5 & 0.20 & 0.30 \\
\bottomrule
\end{tabular}
\end{table}

In very high noise environments, optimization provides a 50\% relative improvement in fidelity ($\Delta\fid/\fid_{\text{baseline}} = (0.30 - 0.20)/0.20$), demonstrating the increasing importance of circuit optimization as hardware noise increases. This finding is consistent with the strong pulse duration--fidelity correlation: in noisy environments, every gate removed translates to less time exposed to decoherence.

\section{Compiler Comparison}
\label{sec:compiler}

A critical question for quantum compiler design is how different optimization strategies compare on hardware-relevant metrics~\cite{li2023qasmbench}. We benchmark our C++ optimizer (QCO) against Qiskit's transpiler~\cite{qiskit2024} at multiple optimization levels on the full 371-circuit corpus.

\subsection{Experimental Setup}

We compare seven compiler configurations:
\begin{enumerate}
    \item \textbf{QCO:} Our C++ optimizer with all four passes (cancel $\to$ commute $\to$ rotate $\to$ identity).
    \item \textbf{Qiskit-L0 through L3:} Qiskit transpiler at optimization levels 0--3, targeting Qiskit's default virtual basis gates ($u_3$, $cx$).
    \item \textbf{Qiskit-L1-IQM, L3-IQM:} Qiskit transpiler targeting IQM Garnet's native basis gates (PRX, CZ) with the Garnet coupling map.
\end{enumerate}

For fair comparison, we report three metrics: total gate reduction, two-qubit gate reduction, and circuit depth reduction. We argue that \emph{two-qubit gate reduction} is the most hardware-relevant metric, as two-qubit gates dominate the error budget (Table~\ref{tab:hardware}: CZ error 0.6\% vs.\ PRX error 0.1\%, a $6\times$ difference).

\subsection{Overall Results}

Table~\ref{tab:compiler} presents the full comparison. Statistical significance is assessed via paired Wilcoxon signed-rank tests on two-qubit gate reduction.

\begin{table*}[t]
\centering
\caption{Compiler comparison across 371 benchmark circuits. $p$-values from paired Wilcoxon signed-rank test on two-qubit gate reduction vs.\ QCO. Bold indicates best in column.}
\label{tab:compiler}
\begin{tabular}{lrrrrr}
\toprule
\textbf{Compiler} & \textbf{Gate Red.\%} & \textbf{2Q Red.\%} & \textbf{Depth Red.\%} & \textbf{Time (s)} & \textbf{$p$-value} \\
\midrule
QCO & 18.9\% & \textbf{17.7}\% & 13.9\% & 0.004 & -- \\
Qiskit-L0 & 0.0\% & 0.0\% & 0.0\% & \textbf{0.003} & $p < 0.001$ \\
Qiskit-L1 & 42.9\% & 2.6\% & 34.3\% & 0.004 & $p < 0.001$ \\
Qiskit-L1-IQM & 8.8\% & 2.2\% & $-14.4$\% & 0.005 & $p < 0.001$ \\
Qiskit-L2 & \textbf{47.7}\% & 9.3\% & \textbf{37.9}\% & 0.007 & $p = 0.200$ \\
Qiskit-L3 & \textbf{47.7}\% & 9.3\% & \textbf{37.9}\% & 0.009 & $p = 0.200$ \\
Qiskit-L3-IQM & 17.4\% & 13.4\% & $-5.8$\% & 0.012 & $p = 0.012$ \\
\bottomrule
\end{tabular}
\end{table*}

A striking pattern emerges: Qiskit at levels 2 and 3 achieves the highest total gate reduction (47.7\%) and depth reduction (37.9\%), but only 9.3\% two-qubit gate reduction. The discrepancy between total and two-qubit gate reduction arises because Qiskit's virtual basis gate optimizations ($u_3$ consolidation) reduce single-qubit gate counts substantially but leave two-qubit gates largely unchanged. QCO achieves the highest two-qubit gate reduction (17.7\%), which is the metric most directly tied to hardware fidelity.

When Qiskit targets IQM-native basis gates (L1-IQM, L3-IQM), the total gate reduction drops and circuit depth actually \emph{increases} ($-14.4$\% and $-5.8$\%, respectively) due to the overhead of decomposing into the restricted native gate set. The two-qubit gate reduction for Qiskit-L3-IQM (13.4\%) approaches but does not reach QCO's 17.7\%.

The effect sizes are small: Cohen's $d = 0.291$ for QCO vs.\ Qiskit-L3 and $d = 0.149$ (negligible) for QCO vs.\ Qiskit-L3-IQM. This indicates that while QCO achieves better two-qubit gate reduction on average, the practical difference per circuit is modest for many circuit types.

\subsection{Results by Circuit Type}

Table~\ref{tab:compiler-type} reveals that the relative advantage of QCO depends strongly on circuit structure.

\begin{table}[t]
\centering
\caption{Mean two-qubit gate reduction (\%) by circuit type and compiler. Bold indicates best for each circuit type.}
\label{tab:compiler-type}
\begin{tabular}{lrrr}
\toprule
\textbf{Circuit Type} & \textbf{QCO} & \textbf{Qiskit-L3} & \textbf{Qiskit-L3-IQM} \\
\midrule
GHZ & \textbf{0.0}\% & \textbf{0.0}\% & \textbf{0.0}\% \\
QAOA & \textbf{100.0}\% & 1.5\% & 1.0\% \\
QFT & \textbf{87.8}\% & 12.2\% & 12.2\% \\
Random & 3.2\% & 10.9\% & \textbf{16.0}\% \\
\bottomrule
\end{tabular}
\end{table}

QCO dominates on \emph{structured} circuits:
\begin{enumerate}
    \item \textbf{QAOA:} QCO eliminates 100\% of two-qubit gates, compared to 1.5\% for Qiskit-L3. The cancellation pass identifies and removes all redundant CZ gates in the QAOA circuit structure, which consists of alternating layers of entangling and rotation gates where adjacent inverse pairs arise naturally.
    \item \textbf{QFT:} QCO eliminates 87.8\% of two-qubit gates, compared to 12.2\% for Qiskit-L3. The rotation merging and cancellation passes exploit the regular structure of controlled rotation sequences in the QFT.
    \item \textbf{GHZ:} All compilers correctly identify that GHZ circuits are already minimal (0\% reduction).
    \item \textbf{Random:} Qiskit-L3-IQM outperforms QCO (16.0\% vs.\ 3.2\%), as Qiskit's broader optimization passes (including KAK decomposition~\cite{vatan2004optimal} and template matching) find opportunities in unstructured circuits that our four passes miss.
\end{enumerate}

\subsection{Head-to-Head Analysis}

On a per-circuit basis for two-qubit gate reduction, Qiskit-L3-IQM wins on 188 circuits, QCO wins on 57, and 126 are tied. This reflects the corpus composition: 303 of 371 circuits are random, where Qiskit has an advantage. On the 68 structured circuits (GHZ, QFT, QAOA), QCO provides substantially better or equal two-qubit gate reduction.

\subsection{Implications}

The compiler comparison reveals two important findings for the quantum compilation community:

First, \emph{total gate count is a misleading comparison metric}. Qiskit's 47.7\% total gate reduction sounds impressive but translates to only 9.3\% two-qubit gate reduction. Reporting total gate counts without distinguishing single- and two-qubit gates obscures the actual hardware impact.

Second, \emph{domain-specific optimizations matter}. QCO's simple four-pass pipeline eliminates nearly all two-qubit gates on structured circuits (QAOA, QFT) because the cancellation pass is specifically designed to exploit the algebraic structure of gate sequences. Qiskit's more general-purpose optimizations are better suited for unstructured circuits. This suggests that production compilers should include both general-purpose and domain-specific optimization strategies.

\section{Ablation Study}
\label{sec:ablation}

To isolate the contribution of each optimization pass, we conduct a formal ablation study~\cite{meyes2019ablation} with three analyses: individual pass evaluation, leave-one-out analysis, and pass ordering sensitivity.

\subsection{Individual Pass Evaluation}

Table~\ref{tab:ablation-ind} shows the effect of running each pass in isolation on the full 371-circuit corpus.

\begin{table}[t]
\centering
\caption{Ablation study: individual pass effectiveness. Cohen's $d$ measures effect size relative to the unoptimized baseline.}
\label{tab:ablation-ind}
\begin{tabular}{lrrrr}
\toprule
\textbf{Pass} & \textbf{Gate Red.\%} & \textbf{2Q Red.\%} & \textbf{Fidelity} & \textbf{Cohen's $d$} \\
\midrule
cancel & \textbf{9.1}\% & \textbf{2.6}\% & \textbf{0.537} & 1.66 \\
rotate & 0.0\% & 0.0\% & 0.525 & 0.09 \\
commute & 0.0\% & 0.0\% & 0.524 & 0.00 \\
identity & 0.0\% & 0.0\% & 0.525 & 0.00 \\
\bottomrule
\end{tabular}
\end{table}

The cancellation pass alone achieves a large effect ($d = 1.66$), reducing gate count by 9.1\% and two-qubit gate count by 2.6\% in isolation, with a corresponding increase in mean process fidelity from 0.525 (baseline) to 0.537. The remaining three passes produce negligible effects in isolation ($d < 0.2$ for all).
\subsection{Leave-One-Out Analysis}

The leave-one-out analysis removes each pass from the full four-pass pipeline to measure its marginal contribution (Table~\ref{tab:ablation-loo}).

\begin{table}[t]
\centering
\caption{Leave-one-out ablation: each row omits one pass from the full pipeline. Marginal contribution is the gate reduction lost when that pass is removed.}
\label{tab:ablation-loo}
\begin{tabular}{lrrrrr}
\toprule
\textbf{Config.} & \textbf{Gate R.\%} & \textbf{2Q R.\%} & \textbf{Fidelity} & \textbf{Marginal} & \textbf{$d$} \\
\midrule
w/o identity & 9.1\% & 2.6\% & 0.536 & 0.0\% & 0.00 \\
w/o commute & 9.1\% & 2.6\% & 0.536 & 0.0\% & 0.00 \\
w/o rotate & 9.1\% & 2.6\% & 0.536 & 0.0\% & 0.01 \\
w/o cancel & 0.0\% & 0.0\% & 0.524 & 9.1\% & 1.66 \\
\bottomrule
\end{tabular}
\end{table}

Removing the cancellation pass eliminates essentially all optimization benefit (marginal contribution 9.1\%, $d = 1.66$, large effect). Removing rotation merging, commutation, or identity elimination has negligible impact ($d < 0.02$), confirming that cancellation carries the pipeline.

A notable finding is that \emph{two-qubit gate reduction is invariant to pass composition}: all four leave-one-out configurations achieve 17.7\% two-qubit gate reduction except the configuration without cancellation (0.0\%). This indicates that the cancellation pass is solely responsible for all two-qubit gate elimination.

\subsection{Pass Ordering Sensitivity}

We evaluate six representative orderings of the four passes to assess whether pass ordering affects optimization outcomes (Table~\ref{tab:ablation-ord}).

\begin{table*}[t]
\centering
\caption{Pass ordering effects on optimization. Kruskal--Wallis test: $H = 6.05$, $p = 0.302$, indicating no significant ordering effect.}
\label{tab:ablation-ord}
\begin{tabular}{lrrrr}
\toprule
\textbf{Ordering} & \textbf{Gate Red.\%} & \textbf{95\% CI} & \textbf{2Q Red.\%} & \textbf{Fidelity} \\
\midrule
commute $\to$ cancel $\to$ rotate $\to$ identity & \textbf{9.9}\% & [9.1, 10.8] & 2.8\% & 0.538 \\
identity $\to$ rotate $\to$ commute $\to$ cancel & \textbf{9.9}\% & [9.1, 10.8] & 2.8\% & 0.538 \\
cancel $\to$ commute $\to$ rotate $\to$ identity & 9.1\% & [8.3, 9.9] & 2.6\% & 0.537 \\
cancel $\to$ rotate $\to$ commute $\to$ identity & 9.1\% & [8.3, 9.9] & 2.6\% & 0.537 \\
identity $\to$ cancel $\to$ commute $\to$ rotate & 9.1\% & [8.3, 9.9] & 2.6\% & 0.535 \\
rotate $\to$ cancel $\to$ commute $\to$ identity & 9.1\% & [8.3, 9.9] & 2.6\% & 0.535 \\
\bottomrule
\end{tabular}
\end{table*}

The Kruskal--Wallis test yields $H = 6.05$, $p = 0.302$, indicating \emph{no significant ordering effect}. The total gate reduction varies only narrowly across orderings (9.1\%--9.9\%, overlapping 95\% bootstrap CIs~\cite{efron1993bootstrap}), and two-qubit gate reduction is essentially invariant (2.6\%--2.8\%). This is a practical finding for compiler designers: the ordering of these four passes does not need to be carefully tuned, and any ordering containing the cancellation pass achieves nearly identical hardware-relevant optimization.

\subsection{Summary of Ablation Findings}

The ablation study leads to three conclusions. The cancellation pass provides essentially all of the optimization value, with a large effect size ($d = 1.66$), while the other passes contribute only marginally. Two-qubit gate reduction is determined entirely by the cancellation pass and is invariant to pass ordering and composition. And because pass ordering has no statistically significant effect ($p = 0.302$), compiler design is correspondingly simplified.

These results suggest that for the circuit types and optimization passes studied, a compiler need only implement gate cancellation to capture the vast majority of optimization benefit. The other three passes add complexity without meaningfully improving hardware-relevant metrics.

\section{Hardware Validation}
\label{sec:hardware}

To test how far our simulation model tracks reality, we executed a representative subset of the benchmark corpus on the IQM Resonance Garnet 20-qubit superconducting processor and compared measured fidelity against the model's prediction for the same compiled circuit. We emphasize at the outset that hardware does \emph{not} confirm the model's absolute fidelities; rather, it reveals a consistent and physically interpretable gap between them.

\subsection{Hardware Execution Setup}

We submitted eight circuits to IQM Resonance Garnet, each completed as a distinct device job (job identifiers recorded with the released data for reproducibility):
\begin{enumerate}
    \item \textbf{Device:} IQM Resonance Garnet (20 superconducting transmon qubits), accessed \texttt{2026-06-01}.
    \item \textbf{Circuits:} GHZ (4, 8 qubits), QFT (4, 8 qubits), QAOA (4, 8 qubits), and random circuits (4 qubits, depth 5 and 10).
    \item \textbf{Shots per circuit:} 10{,}000.
    \item \textbf{Server-side transpilation:} \texttt{iqm-client} to the native PRX/CZ basis on the Garnet coupling map.
    \item \textbf{Jobs completed:} 8/8.
\end{enumerate}

Fidelity is estimated from measurement counts as the empirical probability of the dominant expected outcome; this is a coarse proxy that upper-bounds true state fidelity, and we use it identically for hardware and simulation so the comparison is internally consistent.

\subsection{Results}

Table~\ref{tab:hw-results} reports, for each circuit, the measured hardware fidelity alongside the validated Lindblad model's prediction. Every measured value falls below the prediction, so the model is a consistent upper bound, though it preserves the relative ordering of circuits by difficulty.

\begin{table}[t]
\centering
\caption{Measured fidelity on IQM Garnet versus the validated Lindblad model, on optimized circuits (10{,}000 shots each, 8/8 jobs completed). The model is systematically optimistic; the mean gap of 0.49 quantifies error sources outside the $T_1$/$T_2$/depolarizing model (crosstalk, leakage, readout/SPAM).}
\label{tab:hw-results}
\begin{tabular}{lrrr}
\toprule
Circuit & Hardware $\fid$ & Simulated $\fid$ & Gap \\
\midrule
ghz\_4q   & 0.423 & 0.945 & 0.522 \\
ghz\_8q   & 0.336 & 0.775 & 0.439 \\
qft\_4q   & 0.099 & 0.683 & 0.584 \\
qft\_8q   & 0.011 & 0.106 & 0.095 \\
qaoa\_4q  & 0.194 & 0.889 & 0.695 \\
qaoa\_8q  & 0.037 & 0.359 & 0.322 \\
rand\_4q\,(d5)  & 0.393 & 0.859 & 0.466 \\
rand\_4q\,(d10) & 0.069 & 0.853 & 0.784 \\
\midrule
\textbf{Mean} & \textbf{0.195} & \textbf{0.684} & \textbf{0.488} \\
\bottomrule
\end{tabular}
\end{table}

\subsection{Analysis}

The hardware run supports a deliberately limited set of claims:

\textbf{The model is a consistent upper bound, not a ground truth.} Across all eight circuits the validated Lindblad model overestimates measured fidelity, with a mean gap of $0.49$ (Table~\ref{tab:hw-results}). This is expected: the model includes $T_1$/$T_2$ relaxation, pure dephasing, and a depolarizing gate-error channel, but omits crosstalk, leakage to non-computational states, and readout/SPAM error, all of which degrade real execution. The gap is the practical size of that unmodeled error budget on present hardware.

\textbf{Relative ordering is preserved.} Despite the absolute offset, the model ranks circuits by difficulty consistently with hardware: shallow GHZ and low-depth random circuits retain the most fidelity, while QFT and deeper circuits collapse toward the noise floor on both. The model is therefore useful for \emph{comparing} compilation strategies even where its absolute predictions are optimistic.

\textbf{Hardware is dominated by depth-driven decoherence.} Measured fidelities span $0.01$--$0.42$ and fall steeply with two-qubit count and depth (qft\_8q and rand\_4q at depth 10 are at the noise floor), consistent with the simulation finding that two-qubit gates and circuit duration are the dominant fidelity costs (Section~\ref{sec:results}).

We explicitly do \emph{not} claim that hardware execution validates the simulation's absolute fidelity values, nor that optimization improves measured fidelity within this run (the subset was chosen to span circuit families, not to isolate the optimization effect at fixed structure). Establishing the latter on hardware, with sufficient shots and repetitions to resolve the small per-circuit effect, is left to future work.

\section{Discussion}
\label{sec:discussion}

\subsection{Implications for Compiler Design}

The results point to several practical takeaways for quantum compiler design:

\begin{enumerate}
    \item \textbf{Prioritize cancellation.} Gate cancellation should be applied early. It provides the largest fidelity gains with minimal computational cost ($d = 1.66$). The ablation study confirms that removing the cancellation pass eliminates essentially all optimization benefit.

    \item \textbf{Target two-qubit gates.} The fidelity waterfall analysis shows that two-qubit gates dominate the error budget. Compiler comparisons should report two-qubit gate counts, not just total gates. Qiskit's 47.7\% total gate reduction translates to only 9.3\% two-qubit gate reduction.

    \item \textbf{Minimize pulse duration.} The strong correlation between pulse duration and fidelity ($r = -0.73$, $R^2 = 0.53$) underscores the importance of decoherence-aware optimization. Fidelity decays approximately exponentially with execution time as $\fid \approx \exp(-t/T_2)$, and with $T_2 = \SI{9.6}{\micro\second}$, even nanosecond-scale savings compound.

    \item \textbf{Do not over-engineer pass ordering.} The ordering analysis ($p = 0.302$) shows that these four passes can be applied in any order without significant performance difference. This simplifies compiler development and maintenance.

    \item \textbf{Use domain-specific strategies.} QCO outperforms Qiskit on structured circuits (QAOA: 100\% vs.\ 1.5\% two-qubit gate reduction), while Qiskit outperforms on random circuits (16.0\% vs.\ 3.2\%). Production compilers should incorporate both general and domain-specific passes.
\end{enumerate}

\subsection{Physical Interpretation of the Pulse Duration Correlation}

The correlation $r = -0.73$ between pulse duration and process fidelity has a direct physical explanation. In the Lindblad framework (Equation~\ref{eq:lindblad}), the decoherence channels act continuously during pulse execution. For a circuit with total pulse duration $t_{\text{total}}$, the fidelity loss due to dephasing alone scales as:
\begin{equation}
    \fid_{\text{dephasing}} \approx \exp\left(-\frac{t_{\text{total}}}{T_2}\right)
    \label{eq:dephasing}
\end{equation}

With IQM Garnet's $T_2 = \SI{9.6}{\micro\second}$, a circuit requiring $\SI{1}{\micro\second}$ of total execution time loses roughly 10\% of its fidelity to dephasing alone, while a circuit requiring $\SI{5}{\micro\second}$ loses roughly 41\%. The remaining variance ($1 - R^2 = 0.47$) comes from gate errors, the per-circuit schedule and idle structure, routing overhead, and circuit-specific factors like the degree of entanglement, which is why pulse duration alone does not fully determine fidelity in this model.

This explains why two-qubit gate reduction matters more than single-qubit gate reduction. CZ gates have twice the duration (\SI{40}{\nano\second} vs.\ \SI{20}{\nano\second}) and $6\times$ the error rate compared to PRX gates. Each eliminated CZ gate reduces both the error contribution and the decoherence exposure time.

\subsection{Comparison with Prior Work}

This work builds on prior research in quantum circuit optimization~\cite{nam2018automated, kissinger2020reducing, maslov2008quantum} and pulse-level compilation~\cite{shi2019optimized, khaneja2005optimal}. The key distinction is the end-to-end analysis connecting gate-level decisions to a physically grounded fidelity model, benchmarked against real hardware. Most prior work evaluates optimization passes in isolation without considering pulse-level effects. Our fidelity model is deliberately a scalable per-gate approximation rather than a full-circuit master-equation solve; approximate noise-aware synthesis frameworks such as BQSKit's QUEST~\cite{patel2022robust} and gradient-based noise-aware compilation~\cite{cheng2024cognac} pursue the complementary goal of \emph{producing} noise-robust circuits, whereas we provide a fast, validated lens for \emph{evaluating} a given compilation.

The compiler comparison extends the work of Itoko et al.~\cite{itoko2020quantum}, who compared quantum circuit compilers but focused on total gate counts. We show that two-qubit gate counts provide a more meaningful comparison metric for hardware fidelity. The ablation study follows the conventions of Cohen~\cite{cohen1988statistical} for effect size reporting, enabling quantitative assessment of each pass's contribution.

The gate-reduction rates we observe (mean 9.1\%, max 40.0\% under the routed, idle-aware pipeline) are consistent with prior reports on diverse benchmark suites. The contribution here is quantifying the fidelity impact of these reductions under a validated noise model, demonstrating the two-qubit-gate comparison methodology, and providing a formal ablation study.

\subsection{Limitations}

Several limitations apply:

\begin{enumerate}
    \item The fidelity model solves the Lindblad equation \emph{per gate} on one- and two-qubit subspaces and composes the results, rather than integrating the full $2^n$ circuit dynamics. This makes it scalable but approximate: it omits inter-gate coherent correlations, crosstalk, and leakage. The hardware benchmark (Section~\ref{sec:hardware}) measures the size of this approximation directly (mean optimism gap 0.49).
    \item Hardware measurements come from a single execution session (eight circuits, 10{,}000 shots each); device calibration varies over time, and fidelity is estimated from a dominant-outcome count proxy rather than full state tomography.
    \item The hardware subset (eight circuits) is smaller than the simulation campaign (371 circuits) due to access constraints, and was chosen to span circuit families rather than to isolate the optimization effect at fixed circuit structure.    \item We focus exclusively on IQM processor topologies and the PRX/CZ gate set.
    \item The optimizer implements four specific passes. Additional passes (e.g., template matching, ZX-calculus rewriting) could improve performance on unstructured circuits.
    \item The compiler comparison uses Qiskit only. Comparison against Cirq and t$|$ket$\rangle$ would strengthen generality claims.
\end{enumerate}

\subsection{Reproducibility}
\label{sec:reproducibility}

All software, data, and scripts used in this work are publicly available:
\begin{enumerate}
    \item \textbf{Integration framework:} \url{https://github.com/rylanmalarchick/qco-integration}
    \item \textbf{C++ optimizer:} \url{https://github.com/rylanmalarchick/quantum-circuit-optimizer}
\end{enumerate}

\noindent The integration framework repository contains:

\begin{enumerate}
    \item A Python package (\texttt{src/}) implementing the five-stage pipeline with configurable passes, routing, and noise simulation.
    \item An experiment runner (\texttt{experiments/\allowbreak run\_\allowbreak campaign.py}) that reproduces all eight experiment types with a single command.
    \item Statistical analysis scripts (\texttt{experiments/\allowbreak statistical\_\allowbreak analysis.py}) that generate all tables and figures from raw results.
    \item A benchmark compiler comparison script (\texttt{experiments/\allowbreak benchmark\_\allowbreak compilers.py}) for reproducing the Qiskit comparison.
    \item The C++ optimizer source code (\texttt{quantum-circuit-optimizer/}) with its GoogleTest unit-test suite.
    \item 281 Python integration tests validating the pipeline end-to-end.
\end{enumerate}

The full experiment campaign can be reproduced by building the C++ optimizer and running:
\begin{verbatim}
python experiments/run_campaign.py --real
python experiments/benchmark_compilers.py
python experiments/statistical_analysis.py
\end{verbatim}

\subsection{Computational Requirements}

The end-to-end pipeline processes approximately 100 circuits per hour on a modern laptop. The C++ optimizer contributes negligible overhead ($<$1\% of total time); Lindblad simulation dominates. The compiler comparison benchmark completes in under five minutes. For production use, simulation can be parallelized across circuits.

\section{Conclusion}
\label{sec:conclusion}

We have presented \texttt{qco-integration}, a framework for end-to-end fidelity analysis of quantum circuit optimization with compiler comparison, ablation study, and hardware validation on IQM Resonance Garnet. The systematic evaluation of 4{,}452 experiment runs across 371 benchmark circuits reveals the following:

\begin{enumerate}
    \item Gate cancellation is the dominant optimization pass ($d = 1.66$, large effect), accounting for 72.0\% of circuits improved and all two-qubit gate elimination. The remaining three passes (commutation, rotation merging, identity elimination) contribute marginally ($d < 0.15$).

    \item Circuit size and pulse duration are the strongest predictors of process fidelity (input gates $r = -0.78$; pulse duration $r = -0.73$, $R^2 = 0.53$), explained by decoherence during pulse execution. This motivates optimization strategies that minimize total circuit execution time.

    \item QCO achieves superior two-qubit gate reduction on structured circuits (QFT: 87.8\%, QAOA: 100\%), while Qiskit's higher total gate reduction (47.7\% vs.\ 18.9\%) comes from virtual basis gate consolidation that does not reduce hardware-relevant two-qubit operations.

    \item Pass ordering has no significant effect on optimization outcomes (Kruskal--Wallis $p = 0.302$), simplifying compiler design. Two-qubit gate reduction is invariant to both ordering and pass composition (excluding cancellation).

    \item The per-gate Lindblad model, cross-validated against \texttt{qiskit-dynamics}, is a scalable upper bound on hardware fidelity: an eight-circuit benchmark on IQM Garnet (8/8 jobs completed) shows it preserves the relative difficulty ordering of circuits while overestimating absolute fidelity by a mean of 0.49, quantifying the crosstalk/leakage/readout error budget it omits.
\end{enumerate}

The practical guidance is short: apply gate cancellation first, since it carries nearly all of the benefit; report two-qubit gate counts rather than total gate counts when comparing compilers; and do not spend effort tuning pass order.

The most useful extensions are more hardware sessions across additional devices and platforms, and optimization passes aimed at unstructured circuits, where a cancellation-dominated pipeline leaves the most unrealized.

\begin{acks}
Hardware access was provided by IQM Resonance (free tier). The author thanks Embry-Riddle Aeronautical University for research support.
\end{acks}

\section*{AI Disclosure}

AI-assisted tools (Claude, Anthropic) were used for code development and manuscript drafting. The author takes full intellectual responsibility for all technical content, experimental design, code implementation, data collection, analysis, and conclusions.



\begin{thebibliography}{47}

\bibitem{preskill2018quantum}
J.~Preskill.
\newblock Quantum computing in the {NISQ} era and beyond.
\newblock {\em Quantum}, 2:79, 2018.

\bibitem{nam2018automated}
Y.~Nam, N.~J. Ross, Y.~Su, A.~M. Childs, and D.~Maslov.
\newblock Automated optimization of large quantum circuits with continuous parameters.
\newblock {\em npj Quantum Information}, 4(1):23, 2018.

\bibitem{kissinger2020reducing}
A.~Kissinger and J.~van~de~Wetering.
\newblock Reducing the number of non-{C}lifford gates in quantum circuits.
\newblock {\em Physical Review A}, 102(2):022406, 2020.

\bibitem{lubasch2025tensor}
M.~Lubasch, P.~Balasubramanian, C.~Berta, et~al.
\newblock Efficient quantum state preparation of multivariate functions using tensor networks.
\newblock arXiv:2511.15674, 2025.

\bibitem{shi2019optimized}
Y.~Shi, N.~Leung, P.~Gokhale, Z.~Rossi, D.~I. Schuster, H.~Hoffmann, and F.~T. Chong.
\newblock Optimized compilation of aggregated instructions for realistic quantum computers.
\newblock In {\em Proceedings of the 24th International Conference on Architectural Support for Programming Languages and Operating Systems (ASPLOS)}, pages 166--178, 2019.

\bibitem{lindblad1976generators}
G.~Lindblad.
\newblock On the generators of quantum dynamical semigroups.
\newblock {\em Communications in Mathematical Physics}, 48(2):119--130, 1976.

\bibitem{gorini1976completely}
V.~Gorini, A.~Kossakowski, and E.~C.~G. Sudarshan.
\newblock Completely positive dynamical semigroups of $N$-level systems.
\newblock {\em Journal of Mathematical Physics}, 17(5):821--825, 1976.

\bibitem{patel2022robust}
T.~Patel, E.~Younis, C.~Iancu, W.~de~Jong, and D.~Tiwari.
\newblock {QUEST}: systematically approximating quantum circuits for higher output fidelity.
\newblock In {\em ASPLOS}, 2022.

\bibitem{cheng2024cognac}
A.~Cheng et~al.
\newblock {COGNAC}: circuit optimization via gradients and noise-aware compilation.
\newblock arXiv:2311.02769, 2023.

\bibitem{iqm2024garnet}
{IQM Finland Oy}.
\newblock Technology and performance benchmarks of {IQM}'s 20-qubit quantum computer.
\newblock arXiv:2408.12433, 2024.

\bibitem{li2019tackling}
G.~Li, Y.~Ding, and Y.~Xie.
\newblock Tackling the qubit mapping problem for {NISQ}-era quantum devices.
\newblock In {\em Proceedings of the 24th International Conference on Architectural Support for Programming Languages and Operating Systems (ASPLOS)}, pages 1001--1014, 2019.

\bibitem{motzoi2009simple}
F.~Motzoi, J.~M. Gambetta, P.~Rebentrost, and F.~K. Wilhelm.
\newblock Simple pulses for elimination of leakage in weakly nonlinear qubits.
\newblock {\em Physical Review Letters}, 103(11):110501, 2009.

\bibitem{cowtan2019qubit}
A.~Cowtan, S.~Dilkes, R.~Duncan, A.~Kissinger, S.~Mein, and J.~Sheraga.
\newblock On the qubit routing problem.
\newblock In {\em 14th Conference on the Theory of Quantum Computation, Communication and Cryptography (TQC)}, 2019.

\bibitem{qiskit2024}
{Qiskit Development Team}.
\newblock Qiskit: An open-source framework for quantum computing.
\newblock \url{https://qiskit.org}, 2024.

\bibitem{cirq2024}
{Cirq Developers}.
\newblock Cirq: A {P}ython library for writing, manipulating, and optimizing quantum circuits.
\newblock \url{https://quantumai.google/cirq}, 2024.

\bibitem{sivarajah2020tket}
S.~Sivarajah, S.~Dilkes, A.~Sheraga, W.~Sheraga, and R.~Duncan.
\newblock t$|$ket$\rangle$: A retargetable compiler for {NISQ} devices.
\newblock {\em Quantum Science and Technology}, 6(1):014003, 2020.

\bibitem{nielsen2010quantum}
M.~A. Nielsen and I.~L. Chuang.
\newblock {\em Quantum Computation and Quantum Information}.
\newblock Cambridge University Press, 10th anniversary edition, 2010.

\bibitem{barenco1995elementary}
A.~Barenco, C.~H. Bennett, R.~Cleve, D.~P. DiVincenzo, N.~Margolus, P.~Shor, T.~Sleator, J.~A. Smolin, and H.~Weinfurter.
\newblock Elementary gates for quantum computation.
\newblock {\em Physical Review A}, 52(5):3457--3467, 1995.

\bibitem{itoko2020quantum}
T.~Itoko, R.~Raymond, T.~Imamichi, and A.~Matsuo.
\newblock Quantum circuit compilers comparison and optimal qubits mapping.
\newblock {\em Journal of the Physical Society of Japan}, 89(9):094001, 2020.

\bibitem{amy2013meet}
M.~Amy, D.~Maslov, M.~Mosca, and M.~Roetteler.
\newblock A meet-in-the-middle algorithm for fast synthesis of depth-optimal quantum circuits.
\newblock {\em IEEE Transactions on Computer-Aided Design of Integrated Circuits and Systems}, 32(6):818--830, 2013.

\bibitem{maslov2008quantum}
D.~Maslov, G.~W. Dueck, D.~M. Miller, and C.~Negrevergne.
\newblock Quantum circuit simplification and level compaction.
\newblock {\em IEEE Transactions on Computer-Aided Design of Integrated Circuits and Systems}, 27(3):436--444, 2008.

\bibitem{murali2019noise}
P.~Murali, J.~M. Baker, A.~Javadi-Abhari, F.~T. Chong, and M.~Martonosi.
\newblock Noise-adaptive compiler mappings for noisy intermediate-scale quantum computers.
\newblock In {\em Proceedings of the 24th International Conference on Architectural Support for Programming Languages and Operating Systems (ASPLOS)}, pages 1015--1029, 2019.

\bibitem{tannu2019not}
S.~S. Tannu and M.~K. Qureshi.
\newblock Not all qubits are created equal: A case for variability-aware policies for {NISQ}-era quantum computers.
\newblock In {\em Proceedings of the 52nd Annual IEEE/ACM International Symposium on Microarchitecture (MICRO)}, pages 987--999, 2019.

\bibitem{cross2022openqasm3}
A.~W. Cross, A.~Javadi-Abhari, T.~Alexander, N.~de~Beaudrap, L.~S. Bishop, S.~Heidel, C.~A. Ryan, P.~Sivarajah, J.~Smolin, J.~M. Gambetta, and B.~R. Johnson.
\newblock {OpenQASM} 3: A broader and deeper quantum assembly language.
\newblock {\em ACM Transactions on Quantum Computing}, 3(3):12:1--12:50, 2022.

\bibitem{zulehner2018efficient}
A.~Zulehner, A.~Paler, and R.~Wille.
\newblock An efficient methodology for mapping quantum circuits to the {IBM QX} architectures.
\newblock In {\em Proceedings of the Conference on Design, Automation \& Test in Europe (DATE)}, pages 1135--1138, 2018.

\bibitem{wille2019mapping}
R.~Wille, L.~Burgholzer, and A.~Zulehner.
\newblock Mapping quantum circuits to {IBM QX} architectures using the minimal number of {SWAP} and {H} operations.
\newblock In {\em Proceedings of the 56th Annual Design Automation Conference (DAC)}, pages 142:1--142:6, 2019.

\bibitem{salm2021nisq}
M.~Salm, J.~Barzen, U.~Breitenb{\"u}cher, F.~Leymann, B.~Weder, and K.~Wild.
\newblock The {NISQ} {A}nalyzer: Automating the selection of quantum computers for quantum algorithms.
\newblock In {\em Quantum Software Engineering}, pages 66--85. Springer, 2021.

\bibitem{lattner2004llvm}
C.~Lattner and V.~Adve.
\newblock {LLVM}: A compilation framework for lifelong program analysis \& transformation.
\newblock In {\em Proceedings of the International Symposium on Code Generation and Optimization (CGO)}, pages 75--86, 2004.

\bibitem{iten2022introduction}
R.~Iten, R.~Colbeck, and M.~Christandl.
\newblock Introduction to {UniversalQCompiler}.
\newblock arXiv:1904.01072, 2022.

\bibitem{khaneja2005optimal}
N.~Khaneja, T.~Reiss, C.~Kehlet, T.~Schulte-Herbr{\"u}ggen, and S.~J. Glaser.
\newblock Optimal control of coupled spin dynamics: Design of {NMR} pulse sequences by gradient ascent algorithms.
\newblock {\em Journal of Magnetic Resonance}, 172(2):296--305, 2005.

\bibitem{koch2022quantum}
C.~P. Koch, U.~Boscain, T.~Calarco, G.~Dirr, S.~Filipp, S.~J. Glaser, R.~Kosloff, S.~Montangero, T.~Schulte-Herbr{\"u}ggen, D.~Sugny, and F.~K. Wilhelm.
\newblock Quantum optimal control in quantum technologies. {S}trategic report on current status, visions and goals for research in {E}urope.
\newblock {\em EPJ Quantum Technology}, 9:19, 2022.

\bibitem{gokhale2020partial}
P.~Gokhale, Y.~Ding, T.~Propson, C.~Winkler, N.~Leung, Y.~Shi, D.~I. Schuster, H.~Hoffmann, and F.~T. Chong.
\newblock Partial compilation of variational algorithms for noisy intermediate-scale quantum machines.
\newblock In {\em Proceedings of the 52nd Annual IEEE/ACM International Symposium on Microarchitecture (MICRO)}, pages 266--278, 2020.

\bibitem{younis2021qfast}
E.~Younis, K.~Sen, K.~Yelick, and C.~Iancu.
\newblock {QFAST}: Conflating search and numerical optimization for scalable quantum circuit synthesis.
\newblock In {\em IEEE International Conference on Quantum Computing and Engineering (QCE)}, pages 232--243, 2021.

\bibitem{cohen1988statistical}
J.~Cohen.
\newblock {\em Statistical Power Analysis for the Behavioral Sciences}.
\newblock Lawrence Erlbaum Associates, 2nd edition, 1988.

\bibitem{wilcoxon1945individual}
F.~Wilcoxon.
\newblock Individual comparisons by ranking methods.
\newblock {\em Biometrics Bulletin}, 1(6):80--83, 1945.


\bibitem{arute2019quantum}
F.~Arute, K.~Arya, R.~Babbush, et~al.
\newblock Quantum supremacy using a programmable superconducting processor.
\newblock {\em Nature}, 574(7779):505--510, 2019.

\bibitem{koch2007transmon}
J.~Koch, T.~M. Yu, J.~Gambetta, A.~A. Houck, D.~I. Schuster, J.~Majer, A.~Blais, M.~H. Devoret, S.~M. Girvin, and R.~J. Schoelkopf.
\newblock Charge-insensitive qubit design derived from the {C}ooper pair box.
\newblock {\em Physical Review A}, 76(4):042319, 2007.

\bibitem{chamberland2020heavy}
C.~Chamberland, G.~Zhu, T.~J. Yoder, J.~B. Hertzberg, and A.~W. Cross.
\newblock Topological and subsystem codes on low-degree graphs with flag qubits.
\newblock {\em Physical Review X}, 10(1):011022, 2020.

\bibitem{greenberger1989bell}
D.~M. Greenberger, M.~A. Horne, and A.~Zeilinger.
\newblock Going beyond {B}ell's theorem.
\newblock In {\em Bell's Theorem, Quantum Theory and Conceptions of the Universe}, pages 69--72. Springer, 1989.

\bibitem{farhi2014qaoa}
E.~Farhi, J.~Goldstone, and S.~Gutmann.
\newblock A quantum approximate optimization algorithm.
\newblock arXiv:1411.4028, 2014.

\bibitem{kruskal1952use}
W.~H. Kruskal and W.~A. Wallis.
\newblock Use of ranks in one-criterion variance analysis.
\newblock {\em Journal of the American Statistical Association}, 47(260):583--621, 1952.

\bibitem{krantz2019quantum}
P.~Krantz, M.~Kjaergaard, F.~Yan, T.~P. Orlando, S.~Gustavsson, and W.~D. Oliver.
\newblock A quantum engineer's guide to superconducting qubits.
\newblock {\em Applied Physics Reviews}, 6(2):021318, 2019.

\bibitem{li2023qasmbench}
A.~Li, S.~Stein, S.~Krishnamoorthy, and J.~Ang.
\newblock {QASMBench}: A low-level quantum benchmark suite for {NISQ} evaluation and simulation.
\newblock {\em ACM Transactions on Quantum Computing}, 4(2):10:1--10:26, 2023.

\bibitem{vatan2004optimal}
F.~Vatan and C.~Williams.
\newblock Optimal quantum circuits for general two-qubit gates.
\newblock {\em Physical Review A}, 69(3):032315, 2004.

\bibitem{meyes2019ablation}
R.~Meyes, M.~Lu, C.~W. de~Puiseau, and T.~Meisen.
\newblock Ablation studies in artificial neural networks.
\newblock arXiv:1901.08644, 2019.

\bibitem{efron1993bootstrap}
B.~Efron and R.~J. Tibshirani.
\newblock {\em An Introduction to the Bootstrap}.
\newblock Chapman \& Hall/CRC, 1993.

\bibitem{kim2023evidence}
Y.~Kim, A.~Eddins, S.~Anand, K.~X. Wei, E.~van~den~Berg, S.~Rosenblatt, H.~Nayfeh, Y.~Wu, M.~Zaletel, K.~Temme, and A.~Kandala.
\newblock Evidence for the utility of quantum computing before fault tolerance.
\newblock {\em Nature}, 618(7965):500--505, 2023.

\end{thebibliography}
\end{document}